\documentclass[cis]{ipart_v1}

\firstpage{213}
\Vol{14}
\Issue{4}
\Year{2024}

\usepackage{tikz}
\usetikzlibrary{matrix,decorations.pathreplacing}

\theoremstyle{plain}

\theoremstyle{definition}

\theoremstyle{definition}

\usepackage[pdfpagelabels=true]{hyperref}
\usepackage{mathtools}
\usepackage{amsmath}

\usepackage{amssymb}
\usepackage{latexsym}
\usepackage{booktabs}
\usepackage{caption}
\usepackage{amssymb}

\usepackage{url}

\begin{document}

	\title{PINN-MG: A physics-informed neural network for mesh generation}%
	
	\author[M.~Wang, H.~Li, H.~Zhang, X.~Wu and N.~Li]{Min Wang, Haisheng Li$^\ast$, Haoxuan Zhang, Xiaoqun Wu\blfootnote{* Corresponding author.}
		and Nan Li}

	\begin{abstract}
		In numerical simulation, structured mesh generation often requires a lot of time and manpower investment. The general scheme for structured quad mesh generation is to find a mapping between the computational domain and the physical domain. This mapping can be obtained by solving partial differential equations. However, existing structured mesh generation methods are difficult to ensure both efficiency and mesh quality. In this paper, we propose a structured mesh generation method based on physics-informed neural network, PINN-MG. It takes boundary curves as input and then utilizes an attention network to capture the potential mapping between computational and physical domains, generating structured meshes for the input physical domain. PINN-MG introduces the Navier-Lamé equation in linear elastic as a partial differential equation term in the loss function, ensuring that the neural network conforms to the law of elastic body deformation when optimizing the loss value. The training process of PINN-MG is completely unsupervised and does not require any prior knowledge or datasets, which greatly reduces the previous workload of producing structured mesh datasets. Experimental results show that PINN-MG can generate higher quality structured quad meshes than other methods, and has the advantages of traditional algebraic methods and differential methods.
	\end{abstract}
	
	\maketitle
	
	\section{Introduction}
	\label{sec1}
	
	Mesh generation is the vital part of numerical simulation in different fields, such as computational fluid dynamics \cite{1}, electromagnetics \cite{2}, geophysics \cite{4}, and structural mechanics \cite{3}. According to mesh topology, meshes can be categorized into unstructured and structured types \cite{8}. Unstructured meshes lack regular topological relationships, which reduces the difficulty of generation when dealing with complex geometries. However, they possess complex data structures that result in poorer computational accuracy and robustness \cite{66}. In contrast, structured meshes feature simpler data structures and higher computational efficiency. Despite these advantages, generating high-quality structured meshes remains challenging and often necessitates substantial manual effort in topology design. Additionally, in structured mesh generation, the requirement for the mesh to align orthogonally with the boundary can lead to singular points, significantly affecting numerical results \cite{67}. For example, even minor input errors near these points can cause significant deviations in outputs, thereby reducing the accuracy of simulation results and substantially increasing computational costs.
	
	Traditional structured mesh generation methods are divided into algebraic methods and differential methods \cite{63}. The basic idea of the algebraic method is to transform complex geometric shapes into simple computational domains (regular squares) through a series of coordinate transformations. In this computational space, equidistant and uniform meshes are divided, and then these meshes are mapped to the physical domain (geometric region containing physical boundaries and conditions) to generate meshes. 
	The most commonly used technique in the algebraic method is Transfinite Interpolation (TFI), which was first proposed by Gordon \cite{9}.
	The TFI method keeps the outer boundary stationary and uses the physical coordinates of the boundary points for interpolation to generate mesh. 
	Subsequently, Eriksson \cite{10} adapted this method for mesh generation tasks. Allen \cite{11,12} enhanced the TFI technique by optimizing orthogonality and distribution at the boundary and introduced a smoothing algorithm to address complexities in boundary configurations. 
	Although the above methods are relatively simple, and are suitable for mesh generation of regular geometric bodies, they are less effective when faced with complex boundary domains.
	The simple functions they use are difficult to express the mapping between the computational domain and the physical domain. The generated mesh quality is low, and it is easy to cause uneven distribution problems.
	
	Due to the shortcomings of the algebraic methods, many researchers began to use differential methods to achieve the mesh generation. The differential method regards the mesh generation problem as a boundary value problem \cite{13}, introduces partial differential equations to represent the mapping, and describes the distribution and deformation of mesh points. The solution of Partial Differential Equations (PDEs) requires the input of boundaries as boundary conditions, so structured meshes can be better generated in complex domains. Karman et al. \cite{14} proposed to use the smoothing technology of elliptic partial differential equations to generate mesh, this method can control the mesh point spacing and angles. Huang et al. \cite{15} used partial differential equations to optimize mesh generation, mainly for the complex geometric shapes of rotating machinery. First, a topological structure is established on the plane, and then the initial mesh is generated using TFI and optimized by Laplace and elliptic PDEs to control the mesh distribution and orthogonality. Although the differential method can generate high-quality mesh and is also applicable to complex geometries, the solution process requires high computing resources and time, so the efficiency of large-scale mesh generation is low.
	
	Deep learning has developed rapidly in the context of big data and Graphics Processing Unit (GPU), and the research trend is more inclined towards 3D data \cite{49}, such as 3D model retrieval \cite{48,16}, 3D model completion \cite{17,50} and 3D model generation \cite{46,47,18}. At present, some researchers have reviewed intelligent mesh generation. Lei et al. \cite{19} summarized the current status of intelligent mesh generation, but most of the methods they reviewed belong to computer graphics. This type of mesh is mainly used for rendering and representing the shape of objects, and has no requirements for the mesh quality, but it does not meet the requirements of simulation. In contrast, meshes for industrial simulation have high requirements for mesh quality. The unreasonable angle and distribution will lead to low numerical computation accuracy and non-convergence issues. In the field of industrial simulation, Zhou et al. \cite{20} proposed using convolutional neural networks to identify singular structures and analyze frame fields, construct segmentation streamlines, decompose complex geometric regions into multiple quadrilateral structures, and finally generate high-quality structured meshes. Pan et al. \cite{21} transformed the mesh generation problem into a Markov decision process problem. The intelligent agent observed the state of the environment, performed actions, and received rewards to learn the optimal strategy, automatically generating high-quality meshes. Tong et al. \cite{22} combined the advancing front method with neural networks, selected reference vertices and updated the front boundary through a policy network, iteratively improving mesh quality. However, these methods are all based on supervision, and their dataset construction costs are high and time-consuming.
	
	The Physics-Informed Neural Network (PINN) was first proposed by Raissi et al. \cite{23}. Unlike traditional neural networks \cite{64,65}, PINN can introduce any given physical law described by general nonlinear partial differential equations. Bin et al. \cite{70} proposed using PINN to solve the Eikonal equation by introducing the partial differential equation into the loss function. This is similar to that we set up the loss function when using PINN for mesh generation. 
	In recent years, researchers have proposed mesh generation methods based on PINN. Chen et al. \cite{24} first applied PINN to mesh generation. They regarded the mesh generation task as a mesh optimization task. They first fitted the boundary points by using a decision tree regression model, then designed a neural network model to learn the mapping from the parameter domain to the physical domain, and introduced a dynamic penalty strategy to improve efficiency and convergence. In addition, Chen et al. \cite{25} proposed an improved method to get prior data by introducing auxiliary lines, and designed a composite loss function containing control equation terms, boundary condition terms, and auxiliary line terms. However, the numerical simulation scenarios, which correspond to the partial differential equations used in these methods, are difficult to relate to the process from the computational domain to the physical domain.
	
	Therefore, this paper proposes a novel physics-informed neural network for structured mesh generation via Navier-Lamé equation. The main contributions of this paper are as follows:
	\begin{itemize}  
		\item{We propose PINN-MG, a data-free structured mesh generation method based on physics-informed attention network. By introducing partial differential equation and attention mechanism, the network completes the mesh generation process by capturing the mapping between computational and physical domains.}
		\item{We design a loss function based on the Navier-Lamé equation that is suitable for the structured mesh generation task.} 
		\item{We apply our visualization platform to different meshes to show the effectiveness of our proposed method. Compared to algebraic methods, our proposed method can generate higher quality meshes in a more efficient way.}
	\end{itemize}
	
	The remaining sections of this paper are organized as follows. In \S \ref{sec2}, we introduce background knowledge on linear elasticity problems and their solutions using the Navier-Lamé equation. In \S \ref{sec3}, we provide a new boundary processing method, overall architecture, and implementation details of the proposed physics-informed neural network for structured mesh generation. In \S \ref{sec4}, we introduce and discuss the results on efficiency, mesh quality, and robustness of different test cases. Finally, we conclude the paper and discuss future directions of work in \S \ref{sec5}.
	
	\section{Background}
	\label{sec2}
	In this section, we briefly review the linear elasticity problem and its solution using the Navier-Lamé equation.
	\subsection{Linear elasticity problem}
	\label{sec2.1}
	The linear elasticity problem involves the study of the elastic behavior of materials or structures under external forces, which follows the basic principles of elasticity theory \cite{51}. In engineering and physics, linear elasticity theory is used to describe the relationship between stress and strain in materials under small deformations. The linear elasticity problem under the continuum medium belongs to the case of the small strain assumption, so according to Hooke's Law, the strain $\varepsilon$ should be $0$ when there is no stress. The stress $\sigma$ and strain $\varepsilon$ have a linear relationship, expressed as $\sigma = E\varepsilon$, where $E$ is the Young's modulus. At the same time, we do not need to consider the thermal elastic effects caused by external heat sources.
	\begin{figure}[h]
		\centering
		\includegraphics[scale=.5]{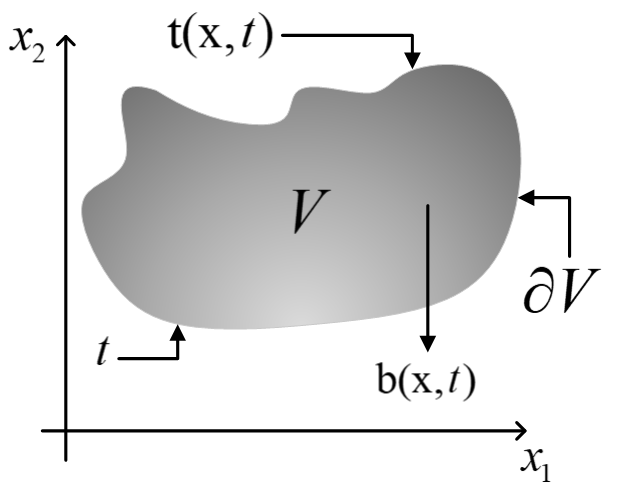}
		\caption{An example of a linear elasticity problem.}
		\label{fig:1}
	\end{figure}
	
	Assuming a linear elastic solid volume $V$ in the 2D space (see Fig. \ref{fig:1}), subjected to volume forces $\mathrm{b}\left (\mathrm{x}, t\right)$ internally and traction forces $\mathrm{t}\left (\mathrm{x}, t\right)$ on the boundary. $\mathrm{b}\left (\mathrm{x}, t\right)$ represents the state of the volume force inside the material at position $\mathrm{x}$ and time $t$. The linear elasticity problem aims to determine displacement $\mathrm{u}\left (\mathrm{x}, t\right)$, strain $\varepsilon \left (\mathrm{x}, t\right)$, and stress $\sigma\left (\mathrm{x}, t\right)$ based on the above boundary and initial conditions. We can describe a linear elasticity object using the Cauchy's equation (momentum conservation equation), constitutive equation, and geometric equation (compatibility equation).
	
	The Cauchy's equation is defined as:
	\begin{equation}
		\nabla \cdot \sigma\left (\mathrm{x}, t\right)+\rho_{0} \mathrm{b}\left (\mathrm{x}, t\right)=\rho_{0} \frac{\partial^{2} \mathrm{u}\left (\mathrm{x}, t\right)}{\partial t^{2}}
		\label{eq:1}
	\end{equation}
	where $\rho_{0}$ represents the initial density of the material. The component form of the Cauchy's equation is:
	\begin{equation}
		\frac{\partial\sigma_{ij}}{\partial x_i}+\rho_0b_j=\rho_0\frac{\partial^2u_j}{\partial t^2}\quad j\in\{1,2\}
		\label{eq:2}
	\end{equation}
	
	The constitutive equation is:
	\begin{equation}
		\sigma(\mathrm{x}, t)=\lambda  \mathrm{T} \mathrm{r}(\varepsilon \left (\mathrm{x}, t\right))I + 2\mu \varepsilon \left (\mathrm{x}, t\right)
		\label{eq:3}
	\end{equation}
	where $\lambda$ and $\mu$ are parameters representing material properties, also known as the Lamé constants. $\mathrm{T} \mathrm{r}(\cdot )$ represents the sum of the main diagonal elements in the square matrix, $I$ represents the unit tensor. The component form of the constitutive equation is:
	\begin{equation}
		\sigma_{ij}=\lambda \delta_{ij} \varepsilon_{u} + 2\mu \varepsilon_{ij}  \quad i,j\in\{1,2\}
		\label{eq:4}
	\end{equation}
	
	The geometric equation is:
	\begin{equation}
		\varepsilon\left (\mathrm{x}, t\right)=\nabla ^{S}\mathrm{u}\left (\mathrm{x}, t\right)=\frac{1}{2}\left (\mathrm{u}\otimes\nabla+\nabla\otimes\mathrm{u}\right)
		\label{eq:5}
	\end{equation}
	where $\nabla ^{S}$ represents the symmetric gradient. The component form of the geometric equation is:
	\begin{equation}
		\varepsilon_{ij}=\frac{1}{2}\left(\frac{\partial u_i}{\partial x_j}+\frac{\partial u_j}{\partial x_i}\right)\quad i,j\in\{1,2\}
		\label{eq:6}
	\end{equation}
	
	For the boundary of a solid $\Gamma\equiv\partial V$, there exist three boundary conditions $\Gamma_{u}$, $\Gamma_{\sigma}$, $\Gamma_{u\sigma}$, and their relationships are as follows:
	\begin{equation}
		\Gamma_u\cup\Gamma_\sigma\cup\Gamma_{u\sigma}=\Gamma \equiv\partial V 
		\label{eq:7}
	\end{equation}
	\begin{equation}
		\Gamma_{u}\cap \Gamma_{\sigma}=\Gamma_{u}\cap\Gamma_{u\sigma}=\Gamma_{u\sigma}\cap\Gamma_{\sigma}=\{\emptyset\}
		\label{eq:8}
	\end{equation}
	where $\Gamma_{u}$ represents the Dirichlet boundary condition, which means that the displacements at the object boundary are all known. $\Gamma_{\sigma}$ represents the Neumann boundary condition, which means that the stress state on the object boundary is known. $\Gamma_{u\sigma}$ represents the mixing condition, which means that the stress state and displacements on the object boundary are both known. Without considering volume forces, only Neumann boundary conditions will cause the solution of PDE to be not unique, because displacement or rotation of the overall rigid body may occur. In contrast, the Dirichlet boundary conditions are usually completely determined, and PDE can find the unique solution. So, we adopt the Dirichlet boundary condition as the boundary condition for the linear elasticity problem.
	\begin{equation}
		\begin{rcases}
			\mathrm{u}\left (\mathrm{x}, t\right)=\mathrm{u}^*\left (\mathrm{x}, t\right)\\
			u_i\left (\mathrm{x}, t\right)={u^*_i}\left (\mathrm{x}, t\right)\quad i\in\{1,2\}
		\end{rcases}
		\quad \forall \mathrm{x}\in \Gamma_u  \quad \forall t
		\label{eq:9}
	\end{equation}
	where $\mathrm{u}^*\left (\mathrm{x}, t\right)$ represents the predefined displacement vector. The initial conditions for the linear elasticity problem are:
	\begin{equation}
		\mathrm{u}\left (\mathrm{x}, 0\right)=0 \quad \forall \mathrm{x}\in V
		\label{eq:10}
	\end{equation}
	
	The structured mesh generation process is a quasi-static problem, because the change of mesh elements in the computational and physical domains with time is very slow during the structured mesh generation process. The quasi-static linear elasticity problem is a linear elasticity problem when the acceleration $\mathrm{a}$ is 0: 
	\begin{equation}
		\mathrm{a} = \frac{\partial^2 \mathrm{u}\left (\mathrm{x}, t\right)}{\partial t^2}\approx 0
		\label{eq:11}
	\end{equation}
	
	The structured mesh generation also does not need to consider volume forces (gravity), so the new Cauchy's equation is:
	\begin{equation}
		\nabla \cdot \sigma \left (\mathrm{x}, t\right)= 0
		\label{eq:12}
	\end{equation}
	
	\subsection{Navier-Lamé equation}
	\label{sec2.2}
	
	The linear elasticity problems can usually be solved by two different methods: the displacement method (Navier-Lamé equation) and the stress method (Beltrami-Michell equation). For the structured mesh generation problem, the essence is to solve the displacement from mesh elements on the computational domain to the vertices on the boundary of the physical domain, so we choose Navier-Lamé equation to solve the relevant PDEs. We need to create an equation that has only displacement as unknown quantity:
	
	\begin{equation}
		\begin{aligned} 
			\nabla \cdot \sigma = \nabla \cdot\left ( \lambda  \mathrm{T} \mathrm{r}(\varepsilon)I +2\mu \varepsilon  \right ) = 0\\ \Longrightarrow \lambda \nabla \cdot \left (\mathrm{T} \mathrm{r}(\varepsilon)I \right ) + 2\mu\nabla \cdot \varepsilon = 0
		\end{aligned}
		\label{eq:13}
	\end{equation}
	
	Then the Navier-Lamé equation can be obtained as:
	\begin{equation}
		\left\{
		\begin{aligned}
			&\left (\lambda + \mu \right )\nabla \left ( \nabla \cdot \mathrm{u} \right )+\mu \nabla^2 \mathrm{u}=0\\
			&\left (\lambda + \mu \right )u_{j,ji} + \mu u_{i,jj}=0 \quad i,j\in\{1,2\}
		\end{aligned}
		\right.
		\label{eq:14}
	\end{equation}
	
	Suppose $\mathrm{u}=(f_x,f_y)$, then the function on each component can be expressed as:
	\begin{equation}
		\left\{
		\begin{aligned}
			&f_x=\mu \left ( \frac{\partial^2g_x}{\partial x^2} + \frac{\partial^2g_x}{\partial y^2}\right ) +\left ( \lambda +\mu  \right )\left ( \frac{\partial^2g_x}{\partial x^2} + \frac{\partial^2g_y}{\partial x \partial y} \right )  \\
			&f_y=\mu \left ( \frac{\partial^2g_y}{\partial x^2} + \frac{\partial^2g_y}{\partial y^2}\right ) +\left ( \lambda +\mu  \right )\left ( \frac{\partial^2g_y}{\partial y^2} + \frac{\partial^2g_x}{\partial x \partial y} \right )  
		\end{aligned}
		\right.
		\label{eq:15}
	\end{equation}
	where $g(x)$ and $g(y)$ are boundary curve functions in the physical domain. Finally, we can get the displacement $\mathrm{u}\left (\mathrm{x}, t\right)$ by the boundary conditions (Eq.\ref{eq:9}), the initial conditions (Eq.\ref{eq:10}), and Eq.\ref{eq:15}. Strain $\varepsilon \left (\mathrm{x}, t\right)$ and stress $\sigma\left (\mathrm{x}, t\right)$ can be obtained through subsequent definitions.
	
	Inspired by Persson et al. \cite{71}, we also treat the structured mesh nodes as truss nodes. By introducing the Navier-Lamé equation to solve for the hydrostatic equilibrium state of the structure under external forces, we determine the final positions of the mesh nodes in the computational domain, thereby generating a structured mesh for the physical domain.

	\begin{figure}[h]
		\centering
		\includegraphics[width=0.75\textwidth]{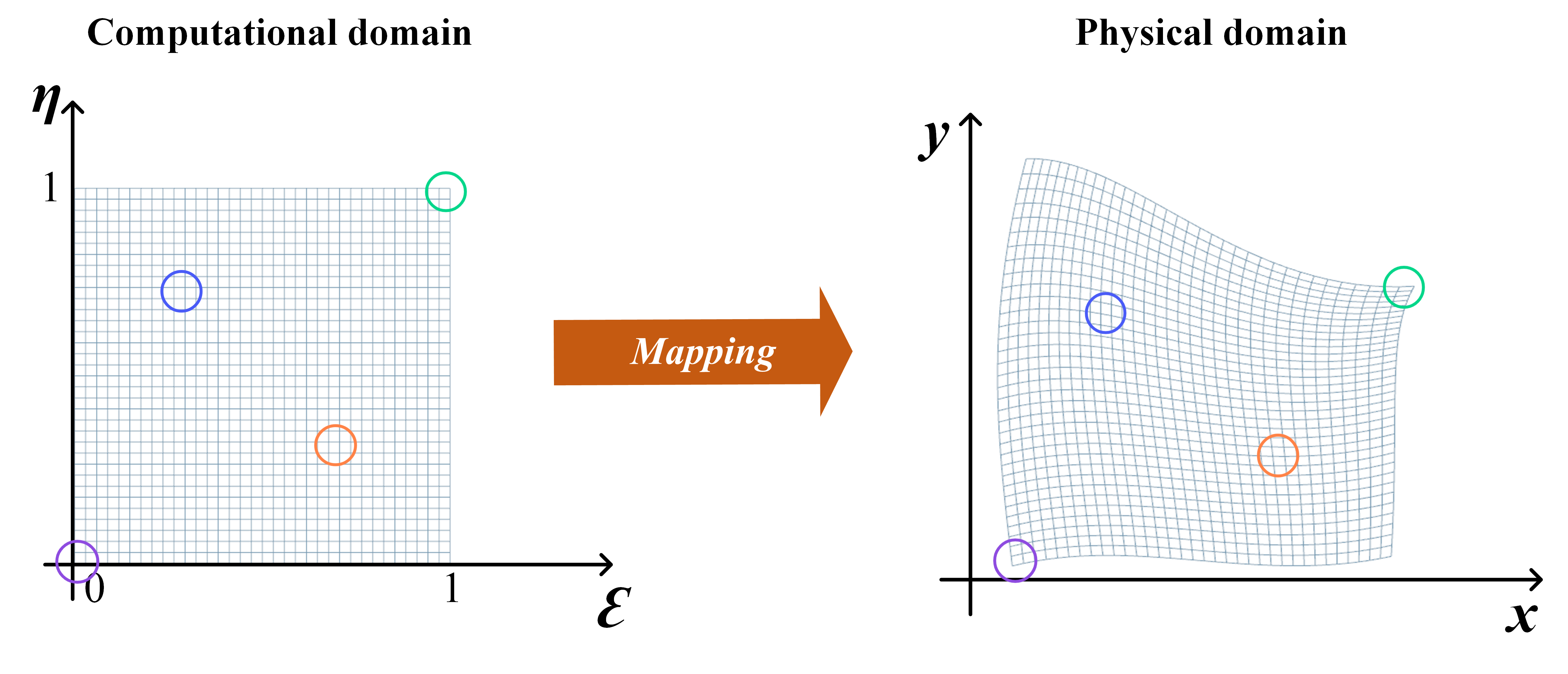}
		\caption{The Mapping relationship between computational domain mesh and physical domain mesh.}
		\label{fig:2}
	\end{figure}

	\section{Proposed approach}
	\label{sec3}
	
\begin{figure}[htbp]
		\centering
		\includegraphics[width=0.98\textwidth]{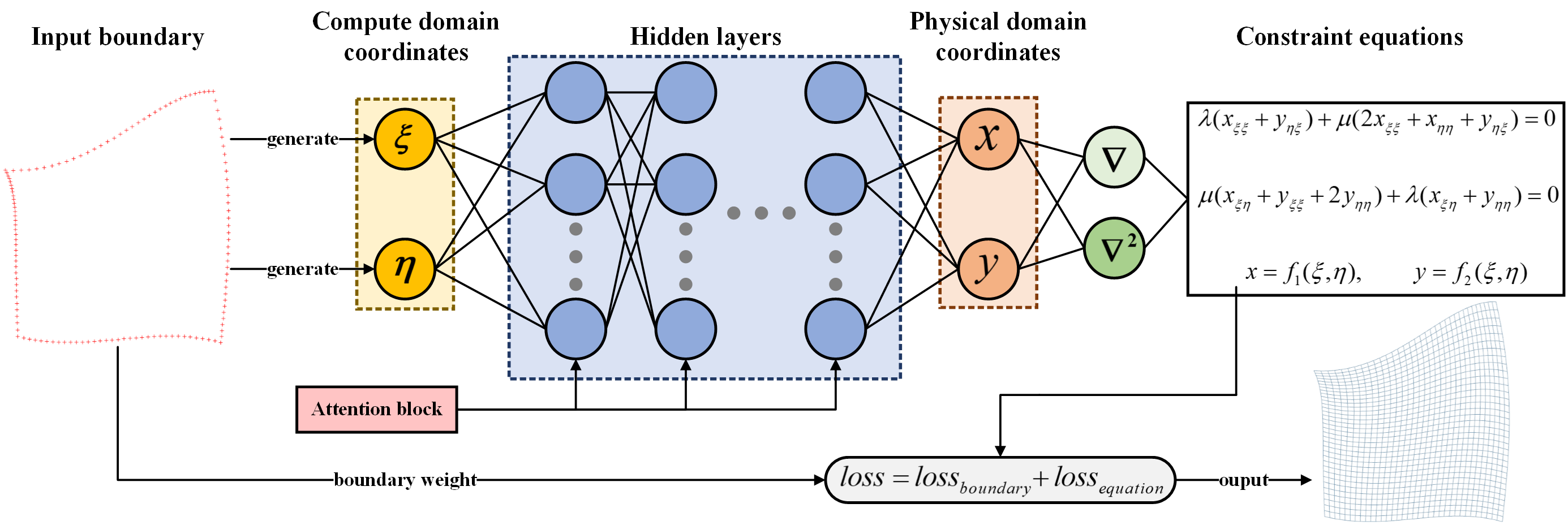}
		\caption{The architecture of the proposed PINN-MG.}
		\label{fig:3}
	\end{figure}

	In this paper, we treat the structured mesh generation task as a mesh deformation problem from the computational domain to the physical domain, as shown in Fig.\ref{fig:2}. We assume that there exists a mapping between the computational domain and the physical domain that can be learned by a neural network. This mapping considers the boundary conditions of the given physical domain (the input boundary curve function) and a series of PDE-based control equations. The mesh deformation process can use the universal approximation theorem of deep neural networks to effectively capture the complex relationship between the computational domain and the physical domain from a high-dimensional nonlinear space. The overall architecture of PINN-MG is shown in Fig.\ref{fig:3}. Structured mesh generation based on PINN-MG includes two main steps: (1) Physical domain boundary condition processing, and (2) Physics-informed neural network construction. We describe these steps in detail in the following sections.
	
	\subsection{Physical domain boundary condition processing}
	\label{sec3.1}

	The PINN-MG we propose is an unsupervised neural network, which means that we do not need any dataset to train our network to generate structured meshes from the input boundary conditions. After inputting the boundary curve, PINN-MG first generates a regular structured mesh of the computational domain, then iteratively trains to find the structured mesh coordinates that satisfy the physical domain boundary conditions. PINN-MG allows the input geometric boundary curves to be expressed as mathematical functions. By matching the number of nodes in the $(\xi, \eta)$ direction of the computational domain mesh, and then sampling the input boundary curve function at the corresponding position, the final boundary conditions of the physical domain are formed.

	However, in some cases, the input geometric boundary is input in the form of control points and cannot form a mathematical function. Therefore, we adopt Support Vector Machine Regression (SVR) models \cite{52} to solve this problem. The basic idea of the SVR model is to find a function in a high-dimensional space that can fit the training data as accurately as possible within a predefined error tolerance. At the same time, we try to maintain the smoothness of the model to prevent overfitting. Different from traditional regression models, SVR models not only try to minimize errors, but also try to reduce model complexity, which can improve the efficiency of PINN-MG. In addition, there are many methods to predict or generate curves from a set of input points, such as random forest regression \cite{53}, decision tree regression \cite{62}, Gaussian process regression \cite{54}, spline regression \cite{55}, AdaBoost regression \cite{56}, and k-neighbor regression \cite{57}. We found through many experiments that the SVR model can more accurately express the boundary conditions of the physical domain, which is crucial for solving PDEs.

	PINN-MG may slightly change some boundary curves to achieve smaller loss values when generating structured meshes. Therefore, we need to add an additional hard boundary condition after generating the structured mesh. Hard boundary conditions ensure that the simulated physical behavior conforms to actual physical constraints at the boundary. First, the values of the boundary points are set to preset mathematical functions (the input physical domain boundary conditions). For points that are not on the boundary, we compute the distance between these points and the boundary points of the hard boundary conditions, and use this distance to perform a weighted average to get the new coordinates of these points $U_{new}$:
	\begin{equation}
		coef=\frac{\left \| \sum_{j=1}^{m} \left ( \left | U_b-U_{Pj}\right |^2 \right ) \right \| }{\sum_{j=1}^{m} \left ( \left | U_b-U_{Pj}\right |^2 \right )}
		\label{eq:16}
	\end{equation}
	\begin{equation}
		U_{new}=U_P + \frac{\sum \left ( U_P-U_{mesh} \right )coef }{\sum coef}
		\label{eq:17}
	\end{equation}
	where $U_P$ represents coordinates of the current boundary condition nodes set, $U_{mesh}$ represents coordinates of mesh nodes set, $U_b$ represents coordinates of the input boundary condition nodes set, $\left \| \cdot  \right \|$ represents the norm of the vector, $\left | \cdot  \right |$ represents the absolute value, $m$ represents the number of boundary points.
	
	\subsection{Physics-informed neural network construction}
	\label{sec3.2}
	
	In this section, we introduce the network architecture of PINN-MG. The architecture of PINN-MG is shown in Fig. 3. The core of this method is to find a suitable mapping between the computational domain and the physical domain from the potential solution space of the underlying PDEs. The network approximates the mapping solution by minimizing the weighted residuals of the physical and boundary constraints, without requiring any prior knowledge or dataset. Different from the control equation used by Chen et al. \cite{24,25}, we use the Navier-Lamé equation as the loss term to constrain the process of the structured mesh generation. The Navier-Lamé equation treats the regular mesh of the computational domain as a linear elastic solid, while the input physical domain boundary conditions are those of the final state solid. This allows the process of structured mesh generation to naturally transform into a mesh deformation task from the computational domain to the physical domain. The loss term based on the Navier-Lamé Equation is back-propagated back into the neural network, which is able to generate a structured mesh in any input boundary curve after training. The final generated mesh nodes are predicted based on the forward feedback of the deep neural network, rather than employing computationally intensive traditional algorithms to solve PDEs.
	
	Inspired by Zhang et al. \cite{43}, PINN-MG uses a deep neural network based on the attention mechanism to predict displacements of structured mesh nodes in the computational domain, thereby generating structured mesh nodes in the physical domain. The network consists of an input layer, a series of hidden layers, attention modules, and an output layer. The input layer first generates the corresponding regular structured mesh in the computational domain based on the vertex coordinates of the physical domain boundary functions, and then samples training points from the interior and boundary curves of the computational domain. After extracting the training points at the input layer, we use a data augmentation method to map the input coordinates into a higher-dimensional space. This enables the neural network to extract more local information from the mesh data, effectively expanding the receptive field of PINN-MG. Therefore, the neural network's input after augmentation $U_{in}$ is:
	\begin{equation}
		U_{in}=U_1 = \left [ \xi, \eta,\tan\left ( \xi, \eta \right ),\cot \left (\xi,\eta\right )\right ] 
		\label{eq:18}
	\end{equation}
	where $\tan\left ( \xi, \eta \right )$ and $\cot \left (\xi,\eta\right )$ represent the tangent and cotangent of the point $\left (\xi,\eta\right )$ respectively.
	
	After the input point coordinates are augmented, we design a series of fully connected neural layers as hidden layers in order to better extract features from them. These hidden layers are activated by a nonlinear activation function, which enables the ability to capture complex mappings. In our previous work \cite{43, 44}, we proposed a structured mesh quality evaluation network based on the dynamic attention mechanism. The dynamic attention mechanism is designed to effectively improve the performance of the model when dealing with structured mesh data by giving the network the ability to focus on the most important parts of the input data. Based on the success of the dynamic attention mechanism, we introduce multiple attention blocks in the current neural network architecture, which are directly connected to the neurons in the network. By dynamically adjusting their weights, the attention blocks are able to respond to changes in the input data in real time, thus enhancing the model’s ability to extract features from the mesh data. This design not only optimizes the information flow of the network, but also enhances the prediction performance of PINN-MG under complex geometric conditions. The nonlinear neural operation of the hidden layer is calculated as:
	\begin{equation}
		U_{i+1}=A_{i+1}\cdot \tanh \left ( W_{i+1}U_{i}+b_{i+1} \right ) \quad i={1,2,\cdots ,l}
		\label{eq:19}
	\end{equation}
	where $W_{i+1}$ and $b_{i+1}$ are the learnable weight parameters and bias tensor of the i-th layer respectively, $l$ is the depth of the hidden layer, $A_{i+1}$ is a weight matrix related to the dynamic attention mechanism, which dynamically adjusts its values in response to changes in the input data. $\tanh(\cdot )$ is the activation function, which can be expressed as:
	\begin{equation}
		\tanh(z)=\frac{e^z-e^{-z}}{e^z+e^{-z}}
		\label{eq:20}
	\end{equation}
	
	The last hidden layer of the neural network is responsible for transmitting high-dimensional features to the output layer,  
	which generates one-dimensional predictions corresponding to point coordinates in the physical domain based on these features. Then using the automatic differentiation function of the PyTorch framework \cite{58}, we can efficiently compute the differential operators of these coordinates. This method is particularly important for us to construct the loss function because it allows us to embed differential operators directly in the loss function, which ensures that the solution output by the network satisfies physical constraints, such as satisfying the underlying PDEs. In physics-informed neural networks, these predictions from the output layer directly relate to subsequent physical tasks, such as structured mesh generation. The loss function in PINN-MG is formulated as follows:
	\begin{equation}
		loss=loss_{boundary}+loss_{equation}
		\label{eq:21}
	\end{equation}
	where $loss_{boundary}$ is the loss term composed of the input boundary curve function, $loss_{equation}$ is the loss term composed of the underlying PDE. According to the control equations selected in this paper and combined with the coordinate system $(\xi, \eta)$, we can get the representation of the Navier-Lamé equation in the structured mesh generation problem:
	\begin{equation}
		\left\{
		\begin{aligned}
			& \lambda \left ( x_{\xi \xi}+y_{\eta \xi} \right ) + \mu \left ( 2x_{\xi \xi}+x_{\eta \eta}+y_{\eta \xi} \right )=0 \\
			& \mu  \left ( x_{\xi \eta}+y_{\xi \xi}+2y_{\eta \eta} \right ) + \lambda  \left ( x_{\xi \eta}+y_{\eta \eta} \right )=0 \\
			& x=f_1 \left (\xi, \eta \right ), \quad y=f_2 \left (\xi, \eta \right )  
		\end{aligned}
		\right.
		\label{eq:22}
	\end{equation}
	
	Then the final loss function is:
	\begin{equation}
		\begin{aligned}
			& loss = \frac{1}{N_1} \sum_{j=1}^{N_1} w_j\\ 
			& \Bigg( \left\| \lambda_a\left( \frac{\partial^2x}{\partial \xi^2} + \frac{\partial^2y}{\partial \xi \partial \eta} \right)  
			+ \mu_a \left( 2\frac{\partial^2x}{\partial \xi^2}+ \frac{\partial^2x}{\partial \eta^2} + \frac{\partial^2x}{\partial y \partial \eta} \right) \right\|^2 \\
			& \quad + \left\| \lambda_a \left( \frac{\partial^2x}{\partial \xi \partial \eta} + \frac{\partial^2y}{\partial \eta^2} \right) 
			+ \mu_a \left( \frac{\partial^2x}{\partial \xi \partial \eta} + \frac{\partial^2y}{\partial \xi^2} + 2\frac{\partial^2y}{\partial \eta^2} \right) \right\|^2 \Bigg) \\
			& \quad + \frac{1}{N_2} \sum_{k=1}^{N_2} w_k \left( \left\| x - f_1(\xi, \eta) \right\|^2 + \left\| y - f_2(\xi, \eta) \right\|^2 \right)
		\end{aligned}
		\label{eq:23}
	\end{equation}	
	
	In order to better enable PINN-MG to focus on important boundaries in the process of generating structured meshes, we add a learnable weight parameter $w_{b_{i}}$ to each boundary curve of the input, where $\sum_{i=1}^{n}{w_{b_{i}}}=1$, and $n$ is the number of boundary curve functions. Therefore, after adding the boundary curves to the loss function of PINN-MG, the neural network can find the most suitable weights for different boundaries during the iterative process of minimizing the loss function. We found that the neural network during the experiment would prefer to add large weights for more complex boundary curves (such as polylines). We believe that it is easier to find problems such as folding and crossing when generating meshes near more complex boundaries, so larger weights will better constrain the generation process of this part of the structured mesh.
	
	Finally, in order to efficiently update and optimize the learnable network parameters, we adopt non-convex optimization algorithms, such as Adaptive Moment Estimation (Adam) algorithm \cite{59} or Stochastic Gradient Descent (SGD) algorithm \cite{60}. The goal of non-convex optimization algorithms is to optimize network weights and bias parameters by iteratively minimizing the loss function. In each iteration, the algorithm computes the loss and its gradient under the current parameters, and then adjusts the parameters based on these gradients, gradually approaching the local minimum of the loss function. This process continues until a certain convergence criterion is reached, such as the gradient change being less than a preset threshold, or after a fixed number of iterations. After converging to the local optimum point, the optimization process stops, and the network parameters at that time are considered to be the best parameter configuration for the given training data and model structure. The Adam algorithm we chose adjusts the learning rate of each parameter automatically, which makes it very effective when training deep neural networks, especially when the parameter space is complex and non-convex.
	
	In addition, the efficiency of PINN-MG is comparable to traditional algebraic-based mesh generation methods, mainly because its feed-forward prediction process mainly involves matrix multiplication operations. After appropriate training, PINN-MG can generate high-quality structured meshes for the input physical domain boundary conditions. Specifically, PINN-MG can be viewed as a high-level regression model that learns the mapping relationship from the computational domain to the physical domain through training. In this process, regular structured mesh node coordinates in the computational domain are used as inputs, and the model finally generates corresponding structured mesh node coordinates in the physical domain by predicting the displacement of the node coordinates. Such a mapping relationship enables PINN-MG to generate adaptable and high quality meshes based on the input boundary conditions and physical laws. At the same time, PINN-MG is able to generate meshes with different resolutions as needed. For example, the thickness of the generated mesh can be easily adjusted by sampling at different densities in the computational domain. For applications that require thick mesh, fewer sample points can be selected; while for simulations that require higher accuracy, the mesh can be refined by increasing the number of sample points.
	
	In summary, PINN-MG is a novel method that combines PINN, traditional mesh generation techniques, and linear elastic theory. This method uses a neural network model that introduces an attention mechanism to achieve intelligent structured meshes generation  by learning the complex mapping relationship between the computational domain and the physical domain. We can further improve the mesh quality and generation efficiency by combining PINN-MG with existing mesh optimization algorithms. For example, the distribution of mesh points can be adjusted by combining the optimization algorithms to meet specific physical or engineering requirements, thus improving the convergence of numerical simulations as well as computational efficiency.
	
	\section{Experiments and results}
	\label{sec4}
	
	In this section, we give the training details of the network and show the experimental results of PINN-MG.
	
	\subsection{Network training procedure}
	\label{sec4.0}
	
	In order to verify the performance of the proposed method in a single-CPU environment, we selected a relatively moderate network size. Specifically, PINN-MG consists of eight hidden layers, with each layer containing 50 neural units. This configuration offers sufficient model complexity and keeps the computational demands within the capabilities of a single-CPU.
	
	\begin{figure}[h]
		\centering
		\includegraphics[scale=.29]{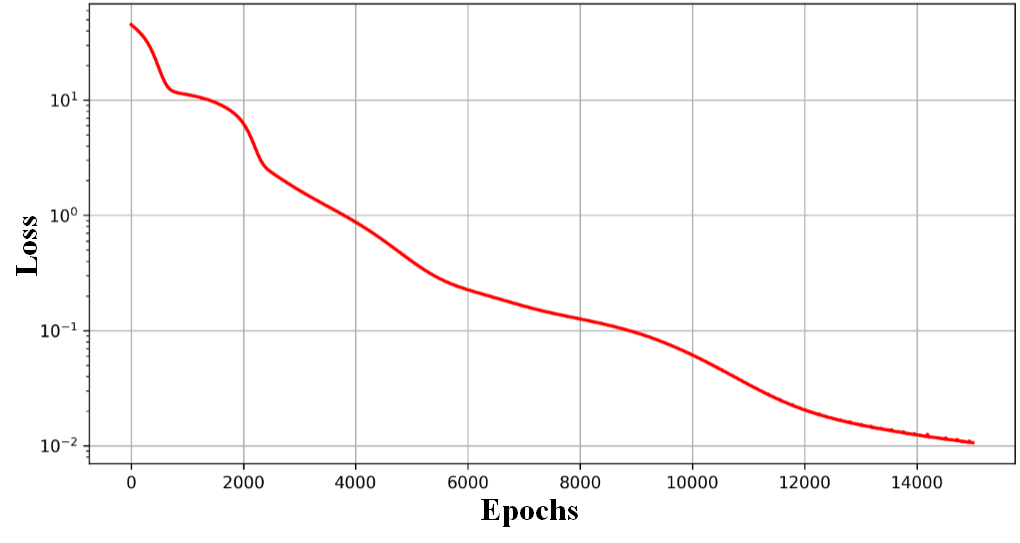}
		\caption{Overall loss of PINN-MG on the structured mesh generation task.}
		\label{fig:4}
	\end{figure}

	During the training process, we apply the Adam optimizer for 15,000 training epochs. We set a small initial learning rate of $1\times 10^{-5}$ to help the model steadily approach the optimal solution during the early stages of training. Additionally, to maintain learning efficiency and avoid falling into the local optimum too early during training, we set a learning rate decay mechanism, and the learning rate decays by 0.99 every 1000 training epochs. The activation function of the last layer of the neural network is Sigmoid, while all other layers use the tanh function. The loss function uses mean squared error loss. Lamé constants $\lambda$ and $\mu$ are set to 1.0 and 0.35 respectively. All test cases are implemented using the PyTorch framework. All visualizations of mesh data in this paper are completed using the MeshLink platform \cite{45}.
	
	\subsection{Comparison studies}
	\label{sec4.1}

	\begin{figure}[h]
		\centering
		\includegraphics[width=0.8\textwidth]{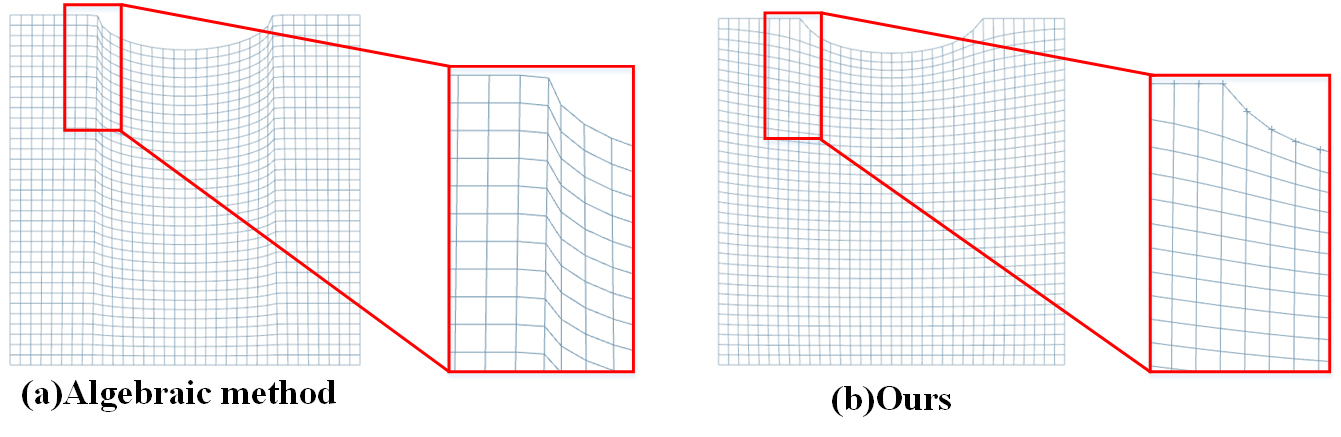}
		\caption{Results of different structured mesh generation methods on model3.}
		\label{fig:7}
	\end{figure}

	\begin{figure}[h]
		\centering
		\includegraphics[width=0.8\textwidth]{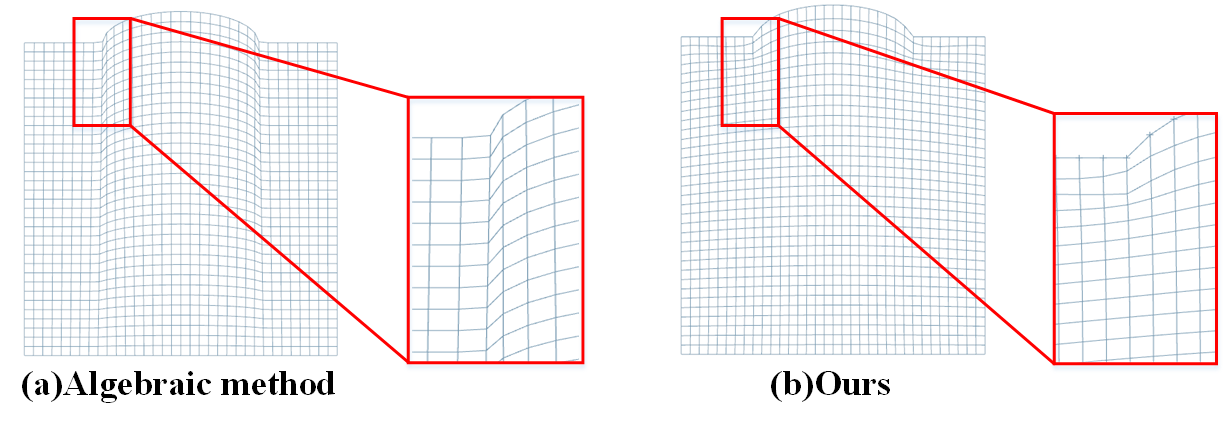}
		\caption{Results of different structured mesh generation methods on model2.}
		\label{fig:6}
	\end{figure}
	
	\begin{figure}[h]
		\centering
		\includegraphics[width=0.8\textwidth]{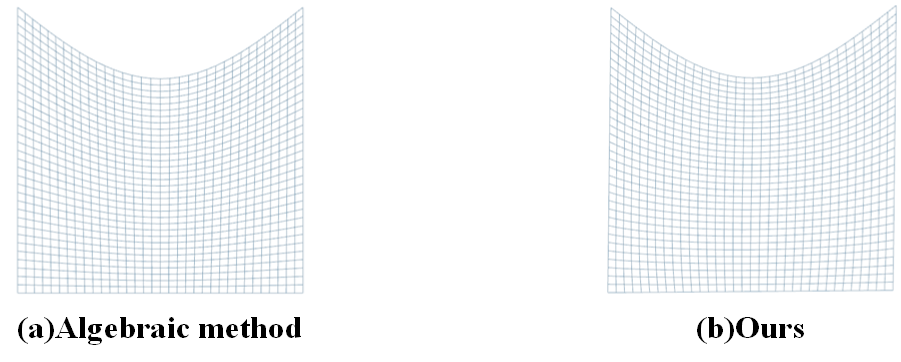}
		\caption{Results of different structured mesh generation methods on model1.}
		\label{fig:5}
	\end{figure}
	
	\begin{figure}[h]
		\centering
		\includegraphics[width=0.8\textwidth]{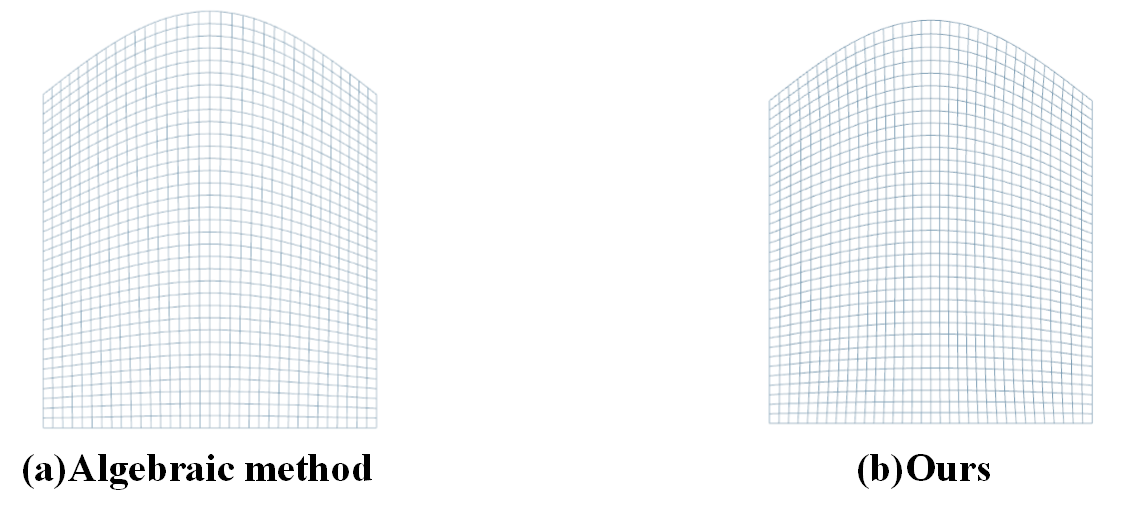}
		\caption{Results of different structured mesh generation methods on model4.}
		\label{fig:8}
	\end{figure}

	\begin{table}[htbp]
		\centering
		\caption{\label{tab1}A comparison of average min/max included angle for different models of meshes. The closer the angle is to 90 the better. The best performance is indicated in bold.}
		\resizebox{\textwidth}{!}{
			\begin{tabular}{llllllll}
				\toprule
				Methods & model1 & model2 & model3& model4 & model5 & model6 & model7\\
				\midrule
				Algebraic& 74.95/105.07 & 83.66/96.40 & 83.48/96.58 & 76.35/103.66 & 83.82/96.20 & 84.14/95.86 & 80.74/99.36\\
				Ours & \textbf{77.76/102.26} & \textbf{85.75/94.21} & \textbf{85.36/94.66} & \textbf{78.25/101.73} & \textbf{85.40/94.62} & \textbf{87.91/92.09} & \textbf{85.60/94.41}\\
				\bottomrule
		\end{tabular}}
	\end{table}

	In this subsection, we compare the performance of our seven models with the traditional algebraic method and the proposed PINN-MG in generating structured meshes under three indicators: average max included angle, mesh generation time and average mesh unit area. The loss convergence of the PINN-MG network model is shown in Fig.\ref{fig:4}. It can be seen that the loss value dropped from more than 10 to close to 0.01, which shows that the training effect of the model is significant.

\begin{figure}[h]
		\centering
		\includegraphics[width=0.8\textwidth]{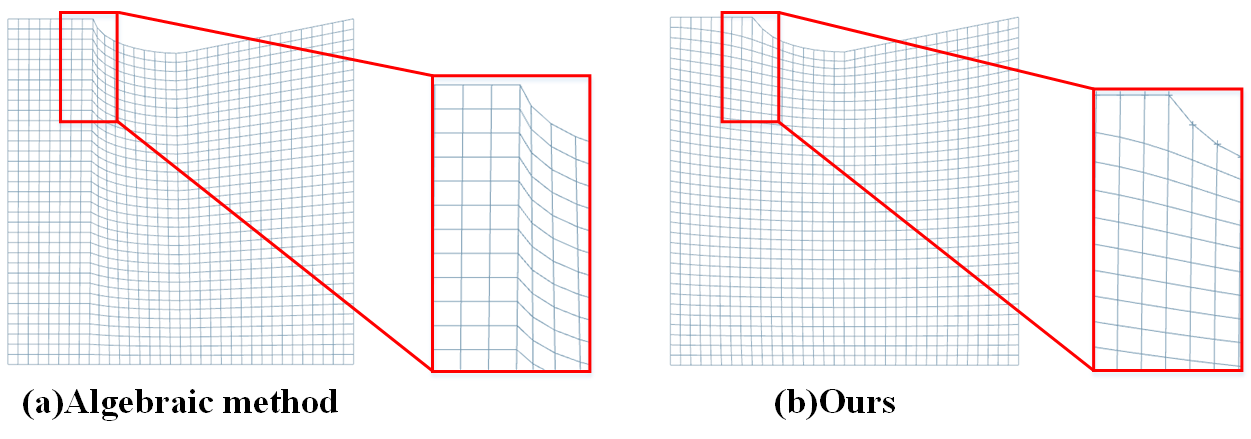}
		\caption{Results of different structured mesh generation methods on model5.}
		\label{fig:9}
	\end{figure}

\begin{figure}[h]
		\centering
		\includegraphics[width=0.8\textwidth]{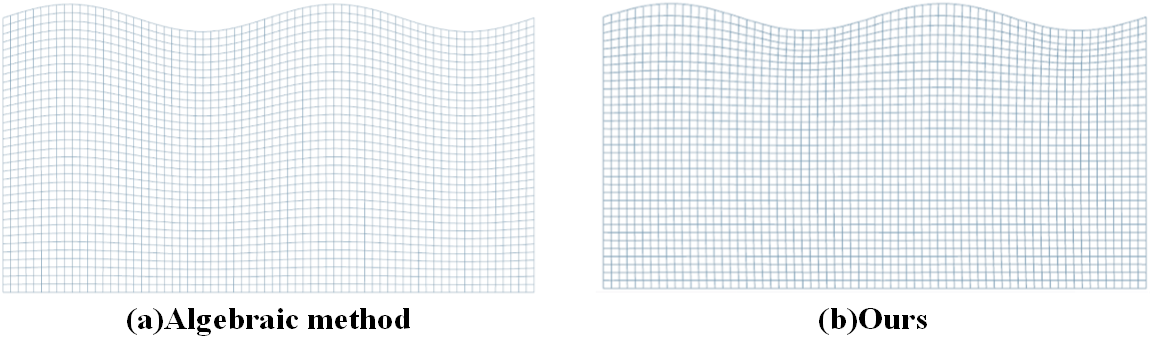}
		\caption{Results of different structured mesh generation methods on model6.}
		\label{fig:10}
	\end{figure}

\begin{figure}[htbp]
		\centering
		\includegraphics[width=0.8\textwidth]{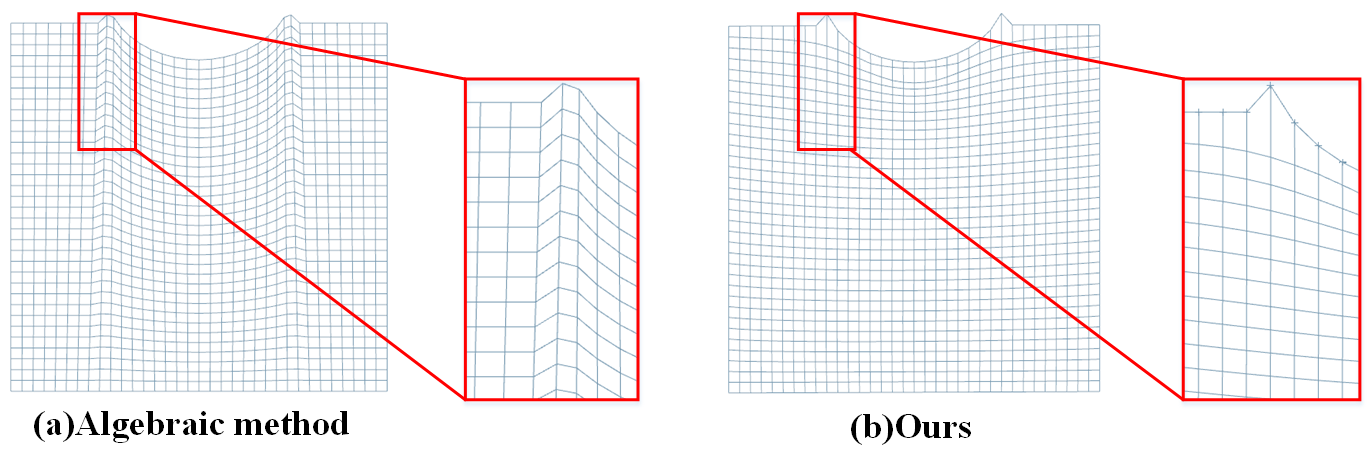}
		\caption{Results of different structured mesh generation methods on model7.}
		\label{fig:11}
	\end{figure}

	We first compared the visual effects of two methods, Fig.\ref{fig:7} to Fig.\ref{fig:11} show the visual results of the algebraic method and PINN-MG for different models in generating structured meshes. As seen in Fig.\ref{fig:5}a, Fig.\ref{fig:8}a, and Fig.\ref{fig:10}a, the mesh generated by the algebraic method is relatively regular, and the mesh in the center region is basically symmetric, but the deformation in the edge region is large. In contrast, Fig.\ref{fig:5}b, Fig.\ref{fig:8}b, and Fig.\ref{fig:10}b show that the mesh generated by PINN-MG is smoother and the mesh deformation in the edge area is smaller than the algebraic method. In Fig.\ref{fig:7}a, Fig.\ref{fig:6}a, Fig.\ref{fig:9}a, and Fig.\ref{fig:11}a, for regions with complex shapes, the algebraic method can result in overlapping mesh units and poor orthogonality in some areas. On the contrary, in Fig.\ref{fig:7}b, Fig.\ref{fig:6}b, Fig.\ref{fig:9}b, and Fig.\ref{fig:11}b, PINN-MG can still generate meshes with good orthogonality in complex regions. The reason our method can generate high-quality meshes under arbitrary shapes is that we not only add partial differential equation terms to the loss function of the network model, but also add boundary constraints, so that the generated mesh can not only conform to the physical laws of linear elastic mechanics, but also fit the boundary as much as possible when generating the mesh. It is also because of the addition of the boundary constraint loss constraint term that when facing complex-shaped regions, even if the curvature of the model changes dramatically, our method can still generate a mesh that meets the simulation calculations. In Table \ref{tab1}, we summarize the average min/max included angle of the mesh generated by the algebraic method and the PINN-MG model for the above 7 models. It can be seen from the experimental results that model6 has the best average min/max included angle, which is 87.91/92.09. Among the seven models, the average min/max included angle of the mesh generated by PINN-MG is all better than that of the algebraic method. In other words, the mesh generated by our proposed method has better performance in terms of shape and orthogonality, and more in line with the expected geometric shape.
	
		\begin{table}[htbp]
		\centering
		\caption{\label{tab2}A comparison of mesh generation time for different models of meshes. Time in seconds.}
		\resizebox{\textwidth}{!}{
			\begin{tabular}{llllllll}
				\toprule
				Methods & model1 & model2 & model3& model4 & model5 & model6 & model7\\
				\midrule
				Algebraic& 0.013 & 0.011 & 0.012 & 0.013 & 0.015 & 0.017 & 0.016\\
				Ours & 0.026 & 0.037 & 0.026 & 0.023 & 0.024 & 0.032 & 0.025\\
				\bottomrule
		\end{tabular}}
	\end{table}
	
	\begin{table}[htbp]
		\centering
		\caption{\label{tab3}A comparison of average mesh unit area($\times 10^{-4}$) for different models of meshes. The average mesh unit area has no concept of relative size, it is determined by mesh density.}
		\resizebox{\textwidth}{!}{
			\begin{tabular}{llllllll}
				\toprule
				Methods & model1 & model2 & model3& model4 & model5 & model6 & model7\\
				\midrule
				Algebraic& 7.276 & 8.988 & 8.313 & 10.024 & 8.265 & 8.525 & 8.399\\
				Ours & 7.275 & 8.992 & 8.309 & 10.026 & 8.264 & 8.651 & 8.401\\
				\bottomrule
		\end{tabular}}
	\end{table}
	
	\begin{table}[htbp]
		\centering
		\caption{\label{tab4}The effect of different activation functions on the proposed PINN-MG.}
		\resizebox{\textwidth}{!}{
			\begin{tabular}{lllllll}
				\toprule
				& sigmod & relu & leakyrelu & tanh(Ours) & elu & selu\\
				\midrule
				Loss value $\downarrow$  & $1.19\times 10$  & $\mathbf{3.30\times 10^{-4}}$ & $3.85\times 10^{-4}$ & $1.06\times 10^{-2}$ & $1.75\times 10^{-2}$ & $1.72\times 10^{-3}$ \\
				average max included angle & 175.92 & 108.48 & 107.42 & \textbf{102.23} & 102.37 & 104.99\\
				Training duration $\downarrow$ & \multicolumn{4}{c}{60-70ms/epoch}  & \multicolumn{2}{c}{200-220ms/epoch} \\
				\bottomrule
		\end{tabular}}
	\end{table}

	In addition to the advantages in mesh quality, our method is similar to the algebraic method in terms of of mesh generation time and average mesh unit area, retaining the benefits of the algebraic method. Table \ref{tab2} lists the mesh generation times for both the algebraic method and our method. The results show that our method also generates meshes very quickly, combining the efficiency of the algebraic method with the introduction of the physical laws of the differential method. At the same time, our method ensures the mesh included angle without losing the mesh unit area, which is verified in Table \ref{tab3}. Table \ref{tab3} lists the average mesh unit area of the meshes generated by the algebraic method and our method. This value is related to the mesh density, so there is no relative size comparison. The results show that the area of the mesh generated by our method is similar to that of the mesh generated by the algebraic method.
	
	In general, our method performs better in mesh generation, especially in complex-shaped region and places with drastic curvature changes. The generated mesh is smoother and more orthogonal, avoiding the problem of excessive deformation that may occur in algebraic methods. Therefore, PINN-MG ensures the quality of generated structured mesh without losing efficiency, providing a more reliable solution for mesh generation in the field of industrial simulation.

	\subsection{Ablation studies}
	\label{sec4.2}
	In this subsection, we conduct an evaluation of the proposed PINN-MG. To verify the effectiveness of the proposed method, we design a series of experiments to explore the performance of different activation functions, control equations, and mesh models in the proposed method. Our experiments focus on three aspects: first, evaluating the impact of different activation functions on model performance; second, analyzing the performance of different control equations in terms of loss convergence; and finally, comparing the performance of different mesh models in generating high-quality meshes.

	\begin{table}[htbp]
		\centering
		\caption{\label{tab5}Loss value of different governing equations. The best performance is indicated in bold.}
		\begin{tabular}{lcccc}
			\toprule
			& Hyperbolic & Monge & Laplace & Navier \\
			&  & -Ampère &  &-Lamé (Ours) \\
			\midrule
			Loss $\downarrow$  & 0.592 & 0.276 & 0.013 & \textbf{0.011}\\
			\bottomrule
		\end{tabular}
	\end{table}
	
	\begin{table}[htbp]
		\centering
		\caption{\label{tab6}A comparison of min/max angle for diferent models of meshes under different governing equations. The closer the min angle and max angle are all to 90 the better. The best performance is indicated in bold.}
		\resizebox{\textwidth}{!}{
			\begin{tabular}{llllllll}
				\toprule
				& model2 & model3 & model5 & model7\\
				\midrule
				Hyperbolic equation& 38.83/139.77 & 40.24/140.47 & 40.590/139.95 & 32.09/148.61 \\
				Monge-Ampère equation& 0.43/178.93 & 1.11/178.02 & -- & 3.39/179.71 \\
				Laplace equation& 40.86/139.24 & 41.01/138.99 & 41.01/138.98 & \textbf{32.61/147.36} \\
				Navier-Lamé equation(Ours) & \textbf{41.77/137.62} & \textbf{41.53/138.33} & \textbf{41.71/138.04} & 32.57/147.50 \\
				\bottomrule
		\end{tabular}}
	\end{table}

	\begin{figure}[h]
		\centering
		\includegraphics[width=0.99\textwidth]{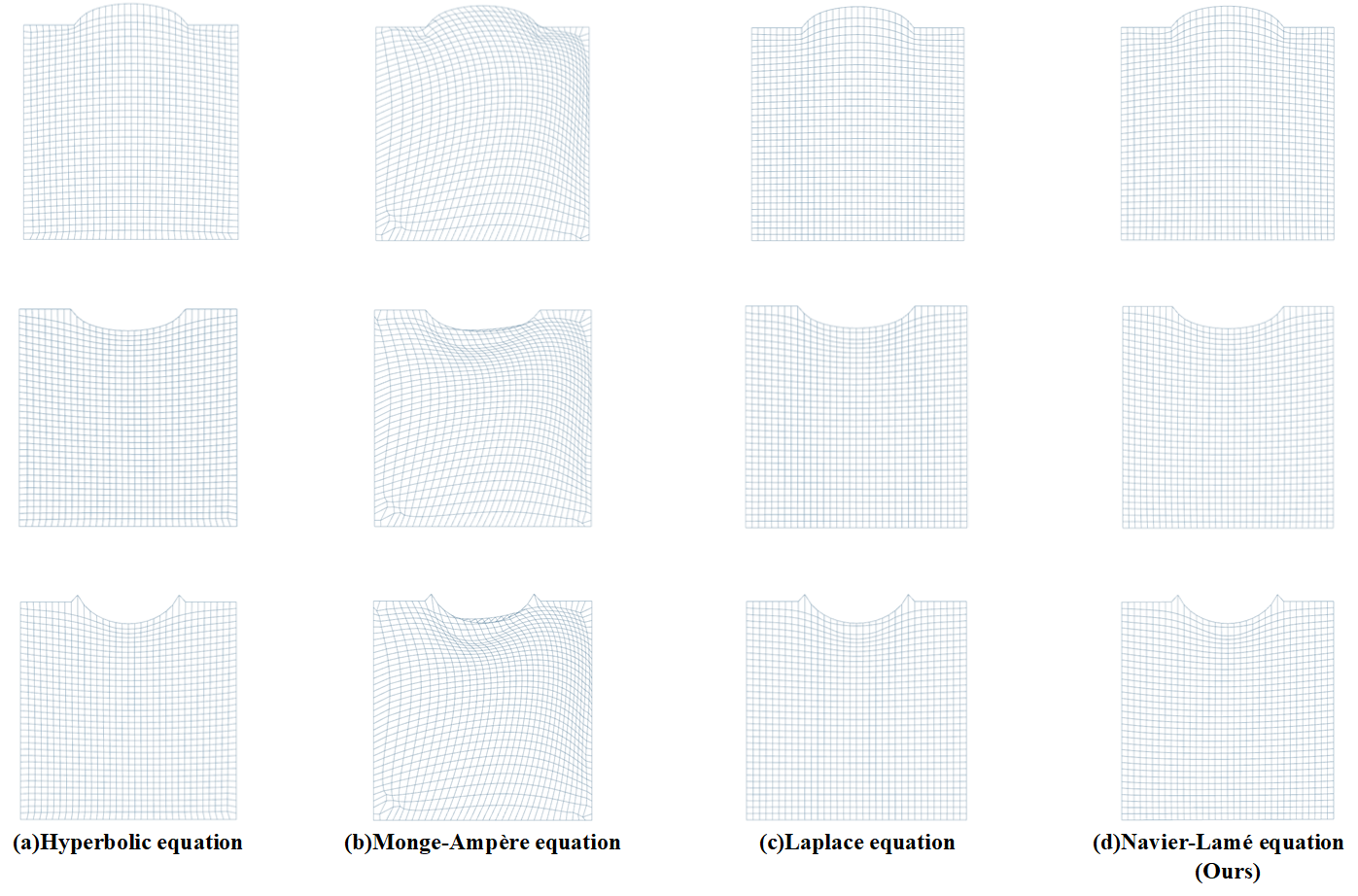}
		\caption{Results of different governing equations on different models.}
		\label{fig:12}
	\end{figure}
	
	In order to study the impact of different activation functions on the performance of PINN-MG, we selected six commonly used activation functions (see Table \ref{tab4}). When selecting the activation function, we focused on three indicators: loss value, average max included angle, and training time. As shown in Table \ref{tab4}, although the relu function performed best in terms of loss value, with a loss value of $3.30\times 10^{-4}$, the quality of the generated mesh was poor, with an average max included angle of only 108.48. This shows that a too low a loss value may cause the mesh vertices to highly fit the laws of partial differential equations, making it difficult to meet the needs of numerical simulation. Since the training time of most activation functions is 60-70 milliseconds per cycle, while the training time of elu and selu is 200 milliseconds per cycle, we chose tanh as our activation function after comprehensive consideration, because tanh performed best in comprehensive indicators, balancing the loss value and mesh quality, and the average max included angle was closest to the ideal 90 degrees, which was 102.23.
	
	A core of PINN is how to effectively integrate physical control equations. To this end, we tested four different governing equations under the PINN-MG framework. These equations have different physical meanings and mathematical properties in numerical simulations, and their selection will significantly impact the convergence speed and accuracy of the network. As shown in Table \ref{tab5}, by comparing the loss values under these control equations, the Navier-Lamé equation performs best in PINN-MG, with the lowest loss value of 0.011, making it easier to converge to a small loss value. This shows that the Navier-Lamé equation has better fitting ability and stability under the proposed PINN-MG framework. This is because the Navier-Lamé equation is mainly used to describe the deformation and stress distribution of elastic materials under the action of external forces. When dealing with mesh generation tasks, the problem can be treated as a deformation process and the mesh can be regarded as an elastic object. its generation process is more in line with elasticity laws and natural elasticity change process. Therefore, we choose the Navier-Lamé equation as the governing equation of our model. Accordingly, Hyperbolic equation (see Eq.\ref{eq:24}) is mainly used to describe wave phenomena, and applied to acoustics, seismology and electromagnetics. Monge-Ampère equation(see Eq.\ref{eq:25}) is used for optical, geometric and optimal transmission problems, applicable to geometric optics and graphics. Laplace equation (see Eq.\ref{eq:26}) is widely used to describe static fields, such as electric fields, gravity fields and pressure distribution in stationary fluids. It can help us understand and predict static field distribution under uniform conditions and is suitable for static field problems and steady-state simulations. 
	
	The Hyperbolic equation is,
	\begin{equation}
		\left\{
		\begin{aligned}
			& x_{\xi}x_{\eta}+y_{\xi}y_{\eta}=0 \\
			&  x_{\xi}y_{\eta}-y_{\xi}x_{\eta}=S \\ 
		\end{aligned}
		\right.
		\label{eq:24}
	\end{equation}
	where $S$ represents the reference area of mesh units.
	
	The Monge-Ampère equation is,
	\begin{equation}
		\left\{
		\begin{aligned}
			& x_{\xi \xi}x_{\eta \eta}-x_{\xi \eta}x_{\eta \xi}-x^2_{\xi}+x^2_{\eta} =0 \\
			& y_{\xi \xi}y_{\eta \eta}-y_{\xi \eta}y_{\eta \xi}-y^2_{\xi}+y^2_{\eta} =0 \\
		\end{aligned}
		\right.
		\label{eq:25}
	\end{equation}
	
	The Laplace equation is,
	\begin{equation}
		\left\{
		\begin{aligned}
			& x_{\xi\xi}\left( x_{\eta}^2 + y_{\eta}^2 \right) - 2 x_{\xi\eta}\left( x_{\xi} x_{\eta} + y_{\xi} x_{\eta} \right) + x_{\eta\eta}\left( x_{\xi}^2 + y_{\xi}^2 \right) &= 0 \\
			& y_{\xi\xi}\left( x_{\eta}^2 + y_{\eta}^2 \right) - 2 y_{\xi\eta}\left( x_{\xi} x_{\eta} + y_{\xi} x_{\eta} \right) + y_{\eta\eta}\left( x_{\xi}^2 + y_{\xi}^2 \right) &= 0
		\end{aligned}
		\right.
		\label{eq:26}
	\end{equation}

	High-quality meshes are critical to the accuracy and stability of numerical simulations, so we further evaluated the performance of different governing equations in generating meshes. As shown in Fig.\ref{fig:12}, the visual results of different models when generating structural meshes under four control equations are shown. We select the min/max angle to compare the mesh quality, these angles ideally should be as close to 90 as possible to ensure uniformity and good geometric properties of the mesh units. As shown in Table \ref{tab6}, we selected multiple mesh models to test the performance of different governing equations under these models. Since the quality of the mesh generated by the Monge-Ampère equation in several other models was poor, we did not continue to do other experimental results under model5. From the table, we can see that the Navier-Lamé equation performs well on model2, model3, and model5, and the min/max angles are either higher or closer than other governing equations, indicating the effectiveness of the Navier-Lamé equation under the PINN-MG model. We can also see that the Monge-Ampère equation is very unsuitable for structured mesh generation tasks. In the case of complex boundary conditions (model 5), PINN-MG based on the Monge-Ampère equation cannot even generate meshes.
	
	\section{Conclusions}
	\label{sec5}
	
	Advances in engineering simulation and artificial intelligence have driven the growing demand for accurate and efficient mesh generation methods. However, there is a lack of research on the intelligent mesh generation process. In this paper, we propose a novel structured mesh generation method based on physical information neural network, PINN-MG, which introduces the Navier-Lamé equation into the loss function, ensuring that the network follows the physical laws of elastic deformation when optimizing the loss value. The training process of PINN-MG does not require any prior knowledge or datasets, which significantly reduces the workload of making structured mesh datasets. Experimental results show that PINN-MG can generate higher quality structured quadrilateral mesh than other methods, and has the advantages of traditional algebraic methods and differential methods. In addition, we also show the visualization effect of PINN-MG on different mesh to verify the effectiveness of our method. In summary, PINN-MG provides a new method for structured mesh generation that is efficient, high-quality, and complies with physical constraints, accelerating the numerical simulation process.
	
	Although our method has achieved good results in improving mesh quality and generation efficiency, it also has certain limitations. For more complex mesh generation tasks, the computational resources and time costs may increase, and when the boundary curve changes, our network requires retraining. Additionally, the chosen network structure and loss function design may not be optimal, and further improvements in efficiency are needed. In the future, we will expand PINN-MG to other types of mesh generation tasks, such as triangular mesh, tetrahedral mesh, and hexahedral mesh. We will further study more efficient neural network architectures that can quickly converge to lower loss values and generate higher quality meshes.
	
	\section*{Acknowledgments}
	This work is supported by Beijing Natural Science Foundation (No. L233026), and National Natural Science Foundation of China (No. 62277001, No. 62272014).

	\address{School of Computer and Artificial Intelligence, Beijing Technology and Business University, Beijing 100048, China\\
		Beijing Key Laboratory of Big Data Technology for Food Safety, Beijing 100048, China\\
		\email{2848608422@qq.com,lihsh@th.btbu.edu.cn,zhxggg613@126.com\\
			wuxiaoqun@btbu.edu.cn,linan@th.btbu.edu.cn}\\
		\received{August 30, 2024}\\
		\accepted{December 12, 2024}\\
	}
	
\let\cleardoublepage\clearpage
	
\end{document}